 \documentclass[iop]{emulateapj}
\usepackage{graphicx}
\usepackage{epstopdf}

\newcommand{\Msol}{\mbox{$M_{\odot}$}}
\newcommand{\msun}{\mbox{$M_{\odot}$}}

\def\deg      {{\ifmmode^\circ\else$^\circ$\fi}} 

 \newcommand{\btxt}[1]{{#1}}
 
 \shorttitle{Weak lensing M -- T relation in COSMOS}
 \shortauthors{Kettula et al.}
 
\begin{document}
 
 \title{Weak lensing calibrated M -- T scaling relation of galaxy groups in the COSMOS field$^{\star}$}
 
\author{K. Kettula\altaffilmark{1},
A. Finoguenov\altaffilmark{1},
R. Massey\altaffilmark{4},
J. Rhodes\altaffilmark{2,3},
H. Hoekstra\altaffilmark{5},
J. E. Taylor\altaffilmark{6},
\btxt{P.F. Spinelli\altaffilmark{7,8},
M. Tanaka\altaffilmark{9}},
O. Ilbert\altaffilmark{10},
P. Capak\altaffilmark{11},
H. J. McCracken\altaffilmark{12},
and A. Koekemoer\altaffilmark{13}
}

%
\altaffiltext{$\star$}{Based on observations with the NASA/ESA {\em
Hubble Space Telescope}, obtained at the Space Telescope Science
Institute, which is operated by AURA Inc, under NASA contract NAS
5-26555; also based on data collected at : the Subaru Telescope, which is operated by
the National Astronomical Observatory of Japan; the XMM-Newton, an ESA science mission with
instruments and contributions directly funded by ESA Member States and NASA; the European Southern Observatory under Large Program 175.A-0839, Chile; Kitt Peak National Observatory, Cerro Tololo Inter-American
Observatory, and the National Optical Astronomy Observatory, which are
operated by the Association of Universities for Research in Astronomy, Inc.
(AURA) under cooperative agreement with the National Science Foundation; 
the National Radio Astronomy Observatory which is a facility of the National Science 
Foundation operated under cooperative agreement by Associated Universities, Inc ; 
and the Canada-France-Hawaii Telescope with MegaPrime/MegaCam operated as a
joint project by the CFHT Corporation, CEA/DAPNIA, the National Research
Council of Canada, the Canadian Astronomy Data Centre, the Centre National
de la Recherche Scientifique de France, TERAPIX and the University of
Hawaii.}  
\altaffiltext{1}{University of Helsinki, Department of Physics, Gustaf H\"allstr\"omin katu 2a, FI-00014 Helsinki, Finland; kimmo.kettula@iki.fi}
\altaffiltext{2}{California Institute of Technology, 1200 East California Boulevard, Pasadena, CA 91125, USA}
\altaffiltext{3}{Jet Propulsion Laboratory, California Institute of Technology, Pasadena, CA 91109, USA}
\altaffiltext{4}{Institute for Computational Cosmology, Durham University, South Road, Durham DH1 3LE, U.K.}
\altaffiltext{5}{Leiden Observatory, Leiden University, Niels Bohrweg 2, NL-2333 CA Leiden, Netherlands}
\altaffiltext{6}{Department of Physics and Astronomy, University of Waterloo, 200 University Avenue West, Waterloo, ON N2L 3G1, Canada}
\altaffiltext{7}{Instituto de Astronomia, Geof\'isica e Ci\^encias Atmosf\'ericas (IAG),Rua do Mat\~ao, 1226 Cidade Universit\'aria 05508-090, S\~ao Paulo, SP - Brazil}
\altaffiltext{8}{Museu de Astronomia e Ci\^encias Afins (MAST), Rua General Bruce, 586 Bairro Imperial de S\~ao Crist\'ov\~ao 20921-030, Rio de Janeiro, RJ - Brazil}
\altaffiltext{9}{National Astronomical Observatory of Japan, Osawa 2-21-1, Mitaka, Tokyo 181-8588, Japan}
\altaffiltext{10}{LAM, CNRS-UNiv Aix-Marseille, 38 rue F. Joliot-Curis, 13013 Marseille, France}
\altaffiltext{11}{Spitzer Science Center, 314-6 Caltech, 1201 E. California Blvd. Pasadena, CA 91125, USA}
\altaffiltext{12}{Institut d'Astrophysique de Paris, UMR 7095, 98 bis Boulevard Arago, 75014 Paris, France}
\altaffiltext{13}{Space Telescope Science Institute, 3700 San Martin Drive, Baltimore, MD 21218, USA}

 \begin{abstract}

\btxt{The scaling between X-ray observables and mass for galaxy clusters and groups is instrumental for cluster based cosmology and an important probe for the thermodynamics of the intracluster gas. We calibrate a scaling relation between the weak lensing mass and X-ray spectroscopic temperature for 10 galaxy groups in the COSMOS field, combined with 55 higher mass clusters from the literature. The COSMOS data includes HST imaging and redshift measurements of 46 source galaxies per arcmin$^2$, enabling us to perform unique weak lensing measurements of low mass systems. Our sample extends the mass range of the lensing calibrated M--T relation an order of magnitude lower than any previous study, resulting in a power-law slope of $1.48^{+0.13}_{-0.09}$. The slope is consistent with the self-similar model, predictions from simulations, and observations of clusters. However, X-ray observations relying on mass measurements derived under the assumption of hydrostatic equilibrium have indicated that masses at group scales are lower than expected. Both simulations and observations suggest that hydrostatic mass measurements can be biased low. Our external weak lensing masses provides the first observational support for hydrostatic mass bias at group level, showing an increasing bias with decreasing temperature and reaching a level of 30--50 \% at 1 keV.
}
\end{abstract}
 
 \keywords{cosmology: observations --  galaxies: groups: general -- gravitational lensing: weak}
 
\setcounter{footnote}{0}

 \section{Introduction}
 \label{sect:intro}

As the largest gravitationally bound objects in the Universe, galaxy clusters and groups have proven to be important cosmological probes. They reside in the high-mass end of the cosmic mass function and have a formation history which is strongly dependent on cosmology. Thus the mass function of galaxy clusters and groups functions as an independent tool for constraining cosmological parameters.

Clusters and groups are now readily detected up to redshifts of unity and above through X-ray emission of hot intracluster gas \footnote{with intracluster gas we refer to the intergalactic gas in both galaxy groups and clusters. \btxt{We follow the convention of referring to those systems with mass lower than $\sim 10^{14}$ \Msol~ as groups and higher as clusters.}}, optical surveys of galaxies and the Sunyaev-Zel'dovich (SZ) effect in the millimeter range. The masses of these systems have typically been inferred through thermal X-ray emission or the velocity dispersion of galaxies. Both of these methods rely on the assumption of hydrostatic or gravitational equilibrium in the cluster or group, which is not always valid. Clusters and groups are found in a myriad of dynamical states and there is increasing evidence for non-thermal pressure support in the intracluster gas, skewing the mass estimates derived under the assumptions of a hydrostatic equilibrium \btxt{\citep[e.g.][]{2007ApJ...668....1N,2008MNRAS.384.1567M,2010ApJ...725.1452S,2012NJPh...14e5018R,2013ApJ...767..116M}}. 

Fortunately, gravitational lensing has proven to be a direct way of measuring cluster and group masses regardless of the dynamical state or non-thermal pressure support in the system. In gravitational lensing the presence of a large foreground mass such \btxt{as} a galaxy cluster or group will bend the light radiating from a background source galaxy. In weak gravitational lensing the ellipticity of a source galaxy is modified, whereas strong lensing \btxt{also} produces multiple images of a single source. The weak lensing induced change in ellipticity is commonly referred to as shear. \btxt{However}, source galaxies typically have a randomly oriented intrinsic ellipticity, \btxt{that} is significantly larger than the lensing induced shear. Therefore the shear has to be averaged over a large sample of source galaxies in order to measure a weak lensing signal used to infer the mass of the lensing system.

The direct mass measurements methods described above are observationally expensive and not always applicable to low mass or high redshift systems. This has spurred the study of mass scaling relations for observables, which can be used as mass proxies. As X-ray observations have proven to be the most efficient way for constructing cluster and group catalogs, typically X-ray observables such as luminosity $L_X$, spectroscopic temperature $T_X$, and thermal energy of the intracluster gas $Y_X = T_X \times M_{gas}$ are used as mass proxies. Consequently, defining and calibrating these X-ray mass proxies is instrumental for cluster and group based cosmology.

The scaling between cluster or group temperature and mass is very fundamental. The simple self-similar model for cluster evolution developed by \citet{1986MNRAS.222..323K}, which assumes pure gravitational heating of intracluster gas, predicts that cluster temperature is a direct measure of the total gravitational potential and thus mass of the system. \btxt{The predicted scaling of mass to temperature is a power-law with a slope of 3/2.} Deviations from the self-similar prediction can consequently be used to study non-gravitational physics affecting the gas.

Unfortunately, cluster and group masses are typically derived from X-ray observations under the assumption of hydrostatic equilibrium \btxt{(HSE)} regardless of dynamical state. Also, temperatures are usually derived from the same observation as hydrostatic masses, introducing possible covariance between the observed quantities. \btxt{The hydrostatic M-T relations typically give power-law slopes in the range of 1.5 -- 1.7 \citep[see][for summaries of recent literature]{2012A&A...539A.120B,2013SSRv..177..247G}. Notably, samples that only include higher-mass systems with temperatures above 3 keV tend to predict M-T relations that have a slope close to the self-similar prediction of 1.5, whereas samples including lower-mass systems tend to predict a slightly steeper proportionality.}

The accuracy of the calibration of mass - temperature scaling can be significantly improved by using independent weak lensing cluster mass measurements. However, this type of studies have only been performed in the cluster mass regime by \citet{2005MNRAS.359..417S,2007MNRAS.379..317H,2010ApJ...721..875O,2011ApJ...737...59J,2012MNRAS.427.1298H,2013ApJ...767..116M}. The aim of this work is to calibrate the scaling between weak lensing masses and X-ray temperatures of the hot intracluster gas for a sample of galaxy groups in the COSMOS survey field. This work is an extension to \citet{2010ApJ...709...97L}, who investigated the scaling between weak lensing mass and X-ray luminosity in the same field.

This paper is organized as follows. We present the data and galaxy group sample used for our analysis in Sections \ref{sect:data} and \ref{sect:sample}, and give details on the X-ray and weak lensing analysis in Sections \ref{sect:X-ray_kt} and \ref{sect:wl_mass}. We present the resulting M -- T relation in Section \ref{sect:scal}, \btxt{discuss our findings in Section \ref{sect:disc}, conclude and summarize our findings in Section \ref{sect:sum}.} Throughout this paper we assume WMAP9 year cosmology \citep{2012arXiv1212.5226H}, with H$_0$ = 70 h$_{70}$ km / s /  Mpc, $\Omega_M$ = 0.28 and $\Omega_{\Lambda}$ = 0.72. All uncertainties are reported at a 68 \% significance, unless stated otherwise.

\section{COSMOS data}
\label{sect:data}

In this Section we briefly present the observations of the COSMOS survey field used for our analysis. The COSMOS survey consists of observations of a contiguous area of 2 square degrees with imaging at wavelengths from radio to X-ray and deep spectroscopic follow-up \citep[see e.g. overview by][]{2007ApJS..172....1S}. 

\subsection{Lensing catalog}
\label{sect:shearcat}

The shear measurements of source galaxies are based on Hubble Space Telescope (HST)  imaging of the COSMOS field using the Advanced Camera for Surveys (ACS) Wide Field Channel (WFC) \citep{2007ApJS..172...38S,2007ApJS..172..196K}. As the COSMOS field was imaged during 640 orbits during HST Cycles 12 and 13, the ACS/WFC imaging of the COSMOS field is the \btxt{HST survey with the largest contiguous area to date}. The derivation of shear measurement is described in detail by \citet{2007ApJS..172..219L}, \citet{2010ApJ...709...97L}, and \citet{2012ApJ...744..159L}. The shear measurement has been calibrated on simulated ACS images containing a known shear \citep{2007ApJS..172..219L}, and we have updated that with each subsequent improvement of the catalog.

The final weak lensing catalog contains accurate shape measurements of 272 538 galaxies, corresponding to approximately 46 galaxies per arcmin$^2$, and a median redshift of z = 1.06. 25 563 of the source galaxies have spectroscopic redshift measurements from the zCOSMOS program \citep{2007ApJS..172...70L}, the remaining source galaxies have photometric redshifts measured \btxt{using} over 30 bands \citep{2009ApJ...690.1236I}.

\subsection{X-ray group catalog}
\label{sect:X-raycat} 
The X-ray group catalog we used has been presented in \citet{2011ApJ...742..125G} and is available online. In brief, we used all XMM-Newton \citep[described in][]{2007ApJS..172...29H,2009A&A...497..635C} and Chandra observations \citep{2009ApJS..184..158E} performed prior to 2010 in catalog construction. Point source removal has been produced separately for Chandra and XMM before combining the data, as described in \citet{2009ApJ...704..564F}, producing a list of 200+ extended sources.  We run a red-sequence finder to identify the galaxy groups following the procedure outlined in \citet{2010MNRAS.403.2063F}. Extensive spectroscopy available for the COSMOS field allowed a 90 \% spectroscopic identification of the z $<$ 1 group sample. \citet{2012ApJ...757....2G} explored the effect of centering by taking an X-ray center or the most massive group galaxy (MMGG).

Previously the X-ray group catalog has been used in \citet{2010ApJ...709...97L} to calibrate the M -- L relation. It has been shown there that there is a correlation between the level of X-ray emission and the significance of the weak lensing signal. In the current work we take advantage of \btxt{the fact} that the significance required to measure the mean X-ray temperature allows us to perform individual mass measurements, and although the sample size is much smaller when compared to the M -- L relation, we do not need to stack several groups in order to produce the results. The high significance of the selected groups also has a much better defined X-ray centering.

\section{Sample selection}
\label{sect:sample}
We selected sources from the COSMOS X-ray group catalog (Section \ref{sect:X-raycat}) with a detection significance of 10$\sigma$ and above. As we choose to exclude cluster cores from temperature determination (see Section \ref{sect:X-ray_kt}) and consequently only use regions with low scatter in pressure \citep{2010A&A...517A..92A}, our sample should be unaffected by selection bias. 

Our initial sample contained 13 sources. However, we excluded the group with id number 6 because X-ray coverage was not sufficient to constrain the spectroscopic temperature. We further excluded the sources with id numbers 246 and 285, as they are located at the edge of the COSMOS field and thus fall outside the coverage of the HST observations (Section \ref{sect:shearcat}). 

The remaining 10 sources in our sample all have a clear X-ray peak with a single optical counterpart and are free of projections \citep[XFLAG = 1]{2007ApJS..172..182F}. \btxt{As our data allows us to extend our lensing analysis out to large radii, possible substructure in the central parts visible in X-rays is not relevant for our mass estimates. Instead, infalling subgroups at cluster outskirts are more important. Based on our X-ray group catalog, we can rule out this kind of substructure at $>$ 20--30 \% level in mass.}

We adopt the coordinates of the X-ray peaks as the locations of the group centers, but we also tested the effect of using the MMGG as a center in performing the lensing analysis (Section \ref{sect:cent}). The properties of the clusters in our sample are presented in Table \ref{tab:cl_prop}. The deep X-ray coverage and high density of background galaxies with determined shear in the COSMOS field allows us to treat each system individually in our analysis. 

\begin{deluxetable}{ccccc}
\tablewidth{0pt}
\tablecaption{Properties of the galaxy group sample.\label{tab:cl_prop}}
 \tablehead{
 \colhead{id \tablenotemark{a}}  & \colhead{N$_{\rm H}$ \tablenotemark{b} } & \colhead{$z$} & \colhead{RA (J2000) \tablenotemark{c}} &  \colhead{Dec (J2000) \tablenotemark{c}} \\ 
 \colhead{ }   & \colhead{[$10^{20}$ cm$^{-2}$]} & \colhead{ }  & \colhead{degrees} & \colhead{degrees} 
 }
\startdata
11 & 1.80 & 0.220 & 150.18980 & 1.65725 \\
17 & 1.78 & 0.372 & 149.96413 & 1.68033 \\
25 & 1.75 & 0.124 & 149.85146 & 1.77319 \\
29 & 1.74 & 0.344 & 150.17996 & 1.76887 \\
120 & 1.80 & 0.834 & 150.50502 & 2.22506 \\
149 & 1.77 & 0.124 & 150.41566 & 2.43020 \\
193 & 1.69 & 0.220 & 150.09093 & 2.39116 \\
220 & 1.71 & 0.729 & 149.92343 & 2.52499 \\
237 & 1.70 & 0.349 & 150.11774 & 2.68425 \\
262 & 1.84 & 0.343 & 149.60007 & 2.82118 \\
\enddata
\tablenotetext{a}{id number in the COSMOS X-ray group catalog (Section \ref{sect:X-raycat})}
\tablenotetext{b}{The LAB weighted average galactic absorption column density \citep{2005A&A...440..775K}}
\tablenotetext{c}{RA and DEC of the X-ray peak.}
\end{deluxetable}

\section{X-ray reduction and analysis}
\label{sect:X-ray_kt}
For the X-ray analysis we used EPIC-pn data from the XMM-Newton wide field survey of the COSMOS field \citep{2007ApJS..172...29H} with the latest calibration information available in October 2012 and XMM Scientific Analysis System (SAS) release xmmsas\_20120621\_1321-12.0.1. We produced event files with the epchain tool and merged the event files of pointings which were within 10 arcmin of the adopted group center for each system. The merged event files were filtered, excluding bad pixels and CCD gaps and periods contaminated by flares, and including only events with patterns 0 -- 4. We generated out-of-time event files, which we subsequently used to subtract events registered during pn readout.
    
We extracted spectra from an annulus corresponding to 0.1 -- 0.5 R$_{500}$ (see Table \ref{tab:x-ray}). \btxt{As differences of a few 10 \% in the inner and outer radii of the X-ray extraction region will be smeared out by the PSF, we determined R$_{500}$ from the virial radius in the X-ray group catalog \citep[Section \ref{sect:X-raycat}, based on the M-L relation of][]{2010ApJ...709...97L}, assuming a halo concentration of 5.} \btxt{The groups were visually inspected  for point sources, which we masked using a circular mask with a 0.5 arcmin radius}. We grouped the spectra to a minimum of 25 counts per bin. 

\btxt{As the groups in the COSMOS field do not fill the FOV, we used the merged event files to extract local background spectra. We selected background regions using the criteria that they are located at a minimum distance of R$_{200}$ ($\sim$ 2--6 arcmin, determined from the X-ray group catalog Section \ref{sect:X-raycat}) and a maximum distance of 10 arcmin from the adopted group center, and that they do not contain any detectable sources. The background spectra where used as Xspec background files in subsequent spectral fits and thus subtracted from the data.}

For X-ray spectroscopy we used an Xspec model consisting of an absorbed thermal APEC component in a 0.5--7.0 keV energy band, with solar abundance tables of \citet{1998SSRv...85..161G} and absorption cross-sections of \citet{1992ApJ...400..699B}. We fixed \btxt{the} metal abundance to 0.3 of the solar value, and used redshift and Galactic absorption column density values listed in Table \ref{tab:cl_prop}. In order to account for spatial variation in the Galactic foreground, we included an additional thermal component with a temperature of 0.26 keV and solar abundance and found that the contribution from this component was negligible.

As the inner radii of the extraction regions is smaller than the EPIC-pn point spread function (PSF), some flux from the excluded central 0.1 R$_{500}$ region might scatter to the extraction region. We accounted for this scatter by extracting spectra from the excluded central regions and fitting them with a similar model as described above. We estimated the scatter to the 0.1--0.5 R$_{500}$ extraction regions using the best-fit model and added the contribution due to the scatter to our analysis. The core regions of groups with id numbers 29 and 220 did not posses a sufficient number of photons to fit a spectrum and we estimate that the scatter from the central region is negligible for these systems. For the remaining systems the fraction of flux in the extraction region scattered from the central region varies between 3 and 21 \% (see Table \ref{tab:x-ray}).

\begin{figure}[tb]
\epsscale{1.0} 
\plotone{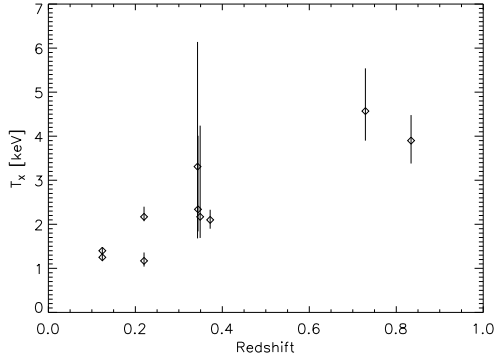}
\caption{\btxt{Plot showing X-ray temperature T$_X$ vs redshift $z$ of the COSMOS systems analyzed in this work}. 
         }
         \label{fig:kTz}
\end{figure}

We detected \btxt{the} thermal emission component in the 0.1 -- 0.5 R$_{500}$ region with a statistical significance of 3.2 -- 24.5$\sigma$ and best-fit temperatures in the range of 1.2 -- 4.6 keV (see  Fig. \ref{fig:kTz} and Table \ref{tab:x-ray}). Thus our sample extends the measurements of weak lensing based M -- T relations to a lower temperature range than previous studies by a factor of four \citep{2007MNRAS.379..317H,2010ApJ...721..875O,2011ApJ...737...59J,2013ApJ...767..116M}. 
  
\begin{deluxetable*}{cccccccc}
\tablewidth{0pt}
\tablecaption{Results of the X-ray analysis.\label{tab:x-ray}} %
\tablehead{\colhead{id} & \colhead{0.1 R$_{500}$ \tablenotemark{a}} & \colhead{0.5 R$_{500}$ \tablenotemark{b}} & \colhead{T$_X$ \tablenotemark{c}} & \colhead{f$_{scat}$ \tablenotemark{d}} & \colhead{sign. \tablenotemark{e}} & \colhead{$\chi^{2}$ \tablenotemark{f}} & \colhead{degrees of} \\ 
\colhead{ } & \colhead{arcmin} & \colhead{arcmin} & \colhead{keV} & \colhead{\%} & \colhead{$\sigma$} & \colhead{ } & \colhead{freedom} 
}
\startdata
11  & 0.35 & 1.77 & 2.2$^{+0.2}_{-0.1}$ & 5       & 24.5 & 273.42 & 263 \\
17  & 0.19 & 0.96 & 2.1$^{+0.2}_{-0.2}$ & 21      & 18.2 & 96.36  & 91  \\
25  & 0.37 & 1.87 & 1.3$^{+0.1}_{-0.1}$ & 3       & 11.8 & 139.40 & 121 \\
29  & 0.18 & 0.89 & 2.3$^{+1.7}_{-0.5}$ & \nodata & 3.2  & 24.75  & 26  \\
120 & 0.13 & 0.67 & 3.9$^{+0.6}_{-0.5}$ & 10      & 16.6 & 66.49  & 69  \\
149 & 0.42 & 2.08 & 1.4$^{+0.1}_{-0.1}$ & 4       & 19.1 & 123.95 & 132 \\
193 & 0.20 & 1.02 & 1.2$^{+0.2}_{-0.1}$ & 14      & 3.9  & 27.54  & 23  \\
220 & 0.16 & 0.79 & 4.6$^{+1.0}_{-0.7}$ & \nodata & 15.8 & 43.49  & 32  \\
237 & 0.20 & 0.99 & 2.2$^{+2.1}_{-0.5}$ & 12      & 5.3  & 24.99  & 26  \\
262 & 0.21 & 1.03 & 3.3$^{+2.8}_{-1.6}$ & 5       & 5.7  & 40.14  & 33  \\
\enddata
\tablenotetext{a}{Inner radii of the extraction region}
\tablenotetext{b}{Outer radii of the extraction region}
\tablenotetext{c}{X-ray temperature of the group}
\tablenotetext{d}{Fraction of the flux in the 0.1 -- 0.5 R$_{500}$ region scattered from the central region}
\tablenotetext{e}{Statistical significance of the thermal X-ray component}
\tablenotetext{f}{$\chi^2$ of the best-fit model}
\end{deluxetable*}

\section{Weak lensing analysis}
\label{sect:wl_mass}

For our weak lensing analysis we used the COSMOS shear catalog. 

\subsection{Lensing signal}
\label{sect:wlsign}
In our analysis we measured the lensing signal independently for each system in our sample in terms of azimuthally averaged surface mass density contrast $\Delta \Sigma(r)$. A spherically symmetric mass distribution is expected to induce a shear, which is oriented tangentially to the radial vector. This signal is also known as the E-mode. The cross-component shear, or B-mode signal, is angled at 45\deg from the tangential shear, and \btxt{the azimuthally averaged value is expected to be consistent with zero for a perfect lensing signal}.

The azimuthally averaged surface mass density contrast is related to the projected tangential shear of source galaxies $\gamma_t$ by
\begin{equation}
\label{eq:dsigma}
\Delta \Sigma(r) = \overline{\Sigma}(< r) - \overline{\Sigma}(r) = \Sigma_{crit} \times \gamma_t(r),
\end{equation}
were $\overline{\Sigma}(< r)$ is the mean surface mass density within the radius $r$, $\overline{\Sigma}(r)$ is the azimuthally averaged surface mass density at radius r, and $\Sigma_{crit}$ is the critical surface mass density. The critical surface mass density depends on the geometry of the lens - source system as
\begin{equation}
\label{eq:sigmacrit}
\Sigma_{crit} = \frac{c^2}{4 \pi G} \frac{D_{OS}}{D_{OL} D_{LS}}.
\end{equation}
Here $c$ is the speed of light, $G$ Newton's gravitational constant and $D_{OS}, D_{OL}$ and $D_{LS}$ the angular diameter distances between observer and source, observer and lens, and lens and source, respectively. 

For each lensing system, we selected the source galaxies from the COSMOS shear catalog with a projected distance of 0.1 -- 4 Mpc in the lens plane and a lower limit for the 68 \% confidence interval for the photometric redshift higher than the redshift of the lensing system. \btxt{Approximately 23 \% of the source galaxies in the lensing catalog have secondary photometric redshift peaks. In order to avoid biasing mass estimates due to catastrophic outliers, we exclude these galaxies from our analysis.
}

\btxt{The lensing signal might be diluted, if a significant number of group galaxies are scattered into the source sample. For instance, \citet{2007MNRAS.379..317H} showed in Fig. 3 that the effect is modest for high mass clusters using  ground based data ($\sim$ 20 \% at R$_{2500}$). As our space based data is deeper, giving a larger number of sources, and we analyse low mass systems with a smaller number of member galaxies, the effect on our sample is significantly smaller. The effect is mainly limited to the central parts of the groups, which we cut out from our analysis. Furthermore, as our photometric redshifts are based on 30+ bands and we exclude source galaxies with secondary redshift peaks, our lensing masses are unaffected by contamination by group members. 
}

\begin{figure}[htb]
\epsscale{1.0} 
\plotone{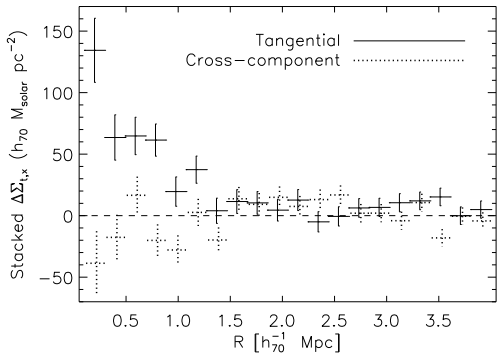}
      \caption{The stacked $\Delta \Sigma$ \btxt{showing the total} tangential (solid data) and cross-component (dotted data) lensing shear of all galaxy groups in the sample. \btxt{Errors represent the total error given by Equation \ref{eq:lens_er}.} The data are binned to 20 equally spaced bins in a radial range of 0.1 to 4 Mpc.
              }
\label{Fig:lenssig}
\end{figure}

We calculated the surface mass density contrast $\Delta \Sigma_{i,j}$ for each lens -- source pair  using Equations \ref{eq:dsigma} and \ref{eq:sigmacrit}. For the computation of $\Delta \Sigma_{i,j}$, spectroscopic redshift was used instead of photometric redshift for those source galaxies where it was available. As we compute $\Delta \Sigma$ at radii greater than 0.1 Mpc, our lensing signals are largely unaffected by non-weak shear or contributions from the central galaxy \citep{2010ApJ...709...97L}. \btxt{As an illustration, we show the combined and binned tangential and cross-component lensing signals for all sources in the sample in Fig. \ref{Fig:lenssig}.} 

The uncertainty of the observed tangential shear $\sigma_{\gamma_t}$ is affected by the measurement error of the shape $\sigma_{meas}$ and the uncertainty due to the intrinsic ellipticity of source galaxies $\sigma_{int}$, known as intrinsic shape noise. \citet{2007ApJS..172..219L, 2010ApJ...709...97L} estimated the intrinsic shape noise of source galaxies in the COSMOS shear catalog to $\sigma_{int}$ = 0.27. 

\btxt{Nearby LSS can also contribute to the uncertainty of lensing mass estimates \citep{2001A&A...370..743H, 2003MNRAS.339.1155H}. For the COSMOS field, \citet{2012MNRAS.420.1384S} found that the LSS affects the shear measurements as an external source of noise,  where the average contribution to the uncertainty of the tangential shear is $\sigma_{LSS} = 0.006$. Thus, the total uncertainty of the tangential shear measurements for each source galaxy can be approximated by:}
\begin{equation}
\label{eq:lens_er}
\sigma_{\gamma_t}^2 \approx \sigma_{meas}^2 + \sigma_{int}^2 + \sigma_{LSS}^2,
\end{equation}
\btxt{since the correlation between the terms  $\sigma_{meas}$ and $\sigma_{LSS}$ is small, the correlation between $\sigma_{int}$ and the other two terms vanishes. For this work we use $\sigma_{meas, j}$ from the updated \citet{2010ApJ...709...97L} catalog, $\sigma_{int}$ = 0.27 and $\sigma_{LSS}$ = 0.006.}

\subsection{Lensing mass estimates}
\label{sect:NFWmass}

Numerical simulations indicate that the density profile of galaxy clusters or groups typically follow the Navarro--Frenk--White (NFW) profile \citep{1997ApJ...490..493N}, given by
\begin{equation}
\label{eq:NFW}
\rho ~(r) = \frac{\delta_c ~\rho_{cr}}{(r / r_s) ~ ( 1 + r / r_s)^{~2}}.
\end{equation}
In this work we define total group mass as the mass inside which the mean NFW mass density $<\rho> = 200~\rho_{cr}$, where $\rho_{cr}$ is the critical density of the Universe at the group redshift $z_d$. We denote this mass by M$_{200}$ and define it as M$_{200} \equiv M(r_{200}) = 200 \rho_{cr} \frac{4}{3} \pi r_{200}^3$. The NFW concentration parameter $c_{200} = r_{200} / r_{s}$ gives the relation between $r_{200}$ and the characteristic scale radius $r_s$. Finally, the density contrast in the NFW profile (Eq \ref{eq:NFW}) is defined as
\begin{equation}
\delta_{c_{200}} = \frac{200}{3} ~\frac{c_{200}^3}{\ln (1 + c_{200}) - \frac{c_{200}}{1 + c_{200}}}.
\end{equation}

\btxt{The analytic solution} for the surface mass density contrast signal corresponding to a NFW profile $\Delta \Sigma_{NFW}$ is given by 
\begin{equation}
\Sigma_{NFW} (x) = \left\lbrace  \begin{array}{lr}
\multicolumn{2}{l}{\frac{2 r_s \delta_c \rho_{cr}}{(x^2 - 1)} \Big[ 1 - \frac{2}{\sqrt{1 - x^2}} {\rm arctanh} \sqrt{\frac{1 - x}{1 + x}} 
~ \Big] } \\
\multicolumn{2}{r}{x < 1,} \\
\frac{2 r_s \delta_c \rho_{cr}}{3} & x = 1, \\
\multicolumn{2}{l}{\frac{2 r_s \delta_c \rho_{cr}}{(x^2 - 1)} \Big[ 1 - \frac{2}{\sqrt{x^2 - 1}} {\rm arctan} \sqrt{\frac{x - 1}{1 + x}}
~ \Big] } \\
\multicolumn{2}{r}{x > 1,}
\end{array} \right.
\end{equation}
where $x = R / r_s$  \citep[e.g.][]{1996A&A...313..697B,2000ApJ...534...34W,2011A&ARv..19...47K}. The solution depends on the mass, concentration parameter and redshift of the lensing system. For this work we assume that M$_{200}$ and $c_{200}$ are related by
\begin{equation}
\label{eq:Mc}
c_{200} = \frac{5.71}{(1 + z_d)^{0.47}} ~ \Bigg( \frac{{\rm M}_{200}}{2.0 \times 10^2 ~h^{-1} {\rm M}_{\odot}} \Bigg)^{-0.084}
\end{equation}
given by \citet{2008MNRAS.390L..64D}. We experimented with letting concentration vary freely, \btxt{however} the shear data did not allow for this extra degree of freedom.  Thus as the redshifts of the systems in our sample are known, the only unknown in the solution of $\Delta \Sigma_{NFW}$ is mass M$_{200}$.

\begin{figure*}[tb]
\epsscale{1.0} 
\plotone{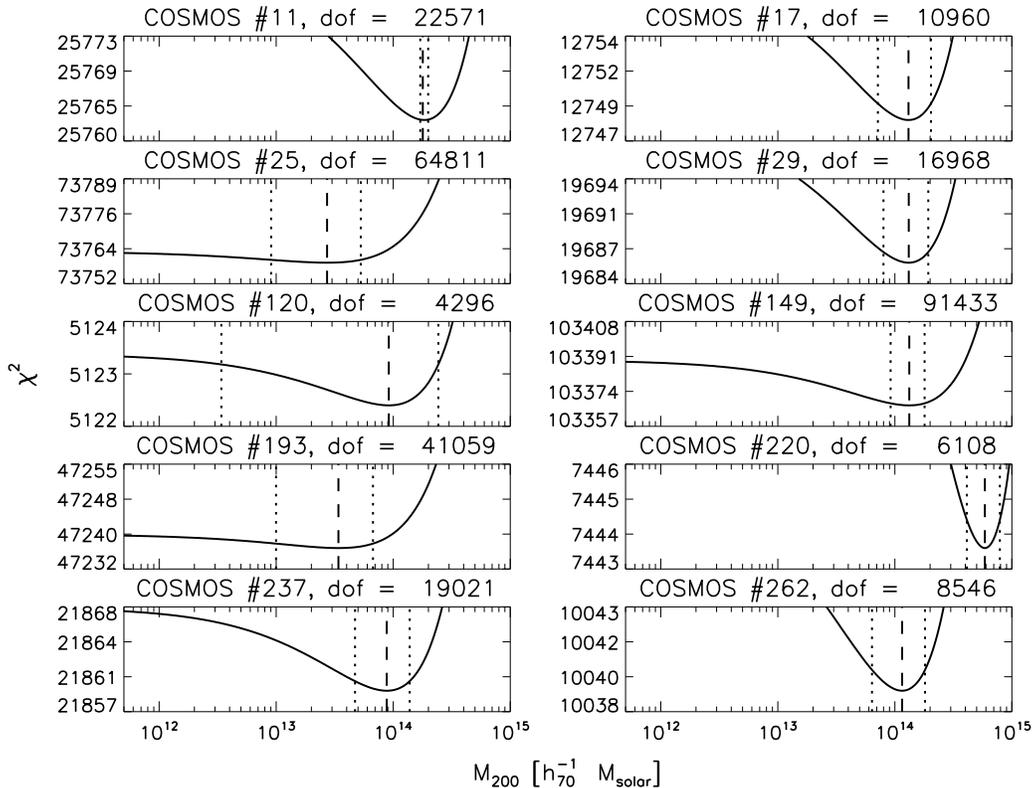}
   \caption{Plot of the $\chi^{2}$ as a function of mass for NFW profile fits to azimuthally averaged mass surface density contrast. The dashed vertical line shows the best-fit M$_{200}$, dotted lines indicate the 1$\sigma$ confidence intervals of M$_{200}$. }
\label{Fig:m200chisq}%
\end{figure*}

\begin{figure*}[htb]
\epsscale{1.} 
\plotone{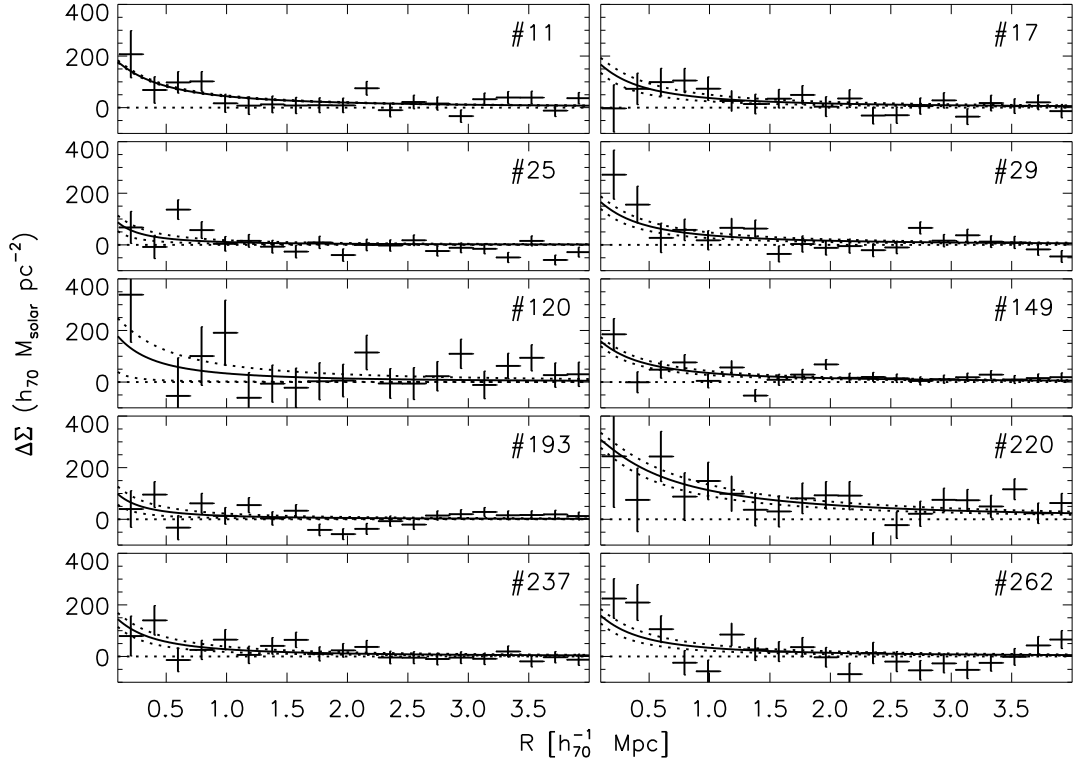}
   \caption{Azimuthally averaged mass surface density contrast $\Delta \Sigma$ profiles of the individual systems used for the weak lensing analysis. The profile is measured in a radial range of 0.1 -- 4 Mpc. Data show the measured $\Delta \Sigma$, solid lines show the $\Delta \Sigma$ of the best-fit NFW density profiles while the dotted lines indicate the statistical uncertainty of the fitted profiles. \btxt{The profile fits are performed to un-binned data, here the data are binned to 20 equally spaced radial bins for plot clarity.}
   			}
\label{Fig:m200prof}
\end{figure*}

\begin{figure}[htb]
\epsscale{1.0} 
\plotone{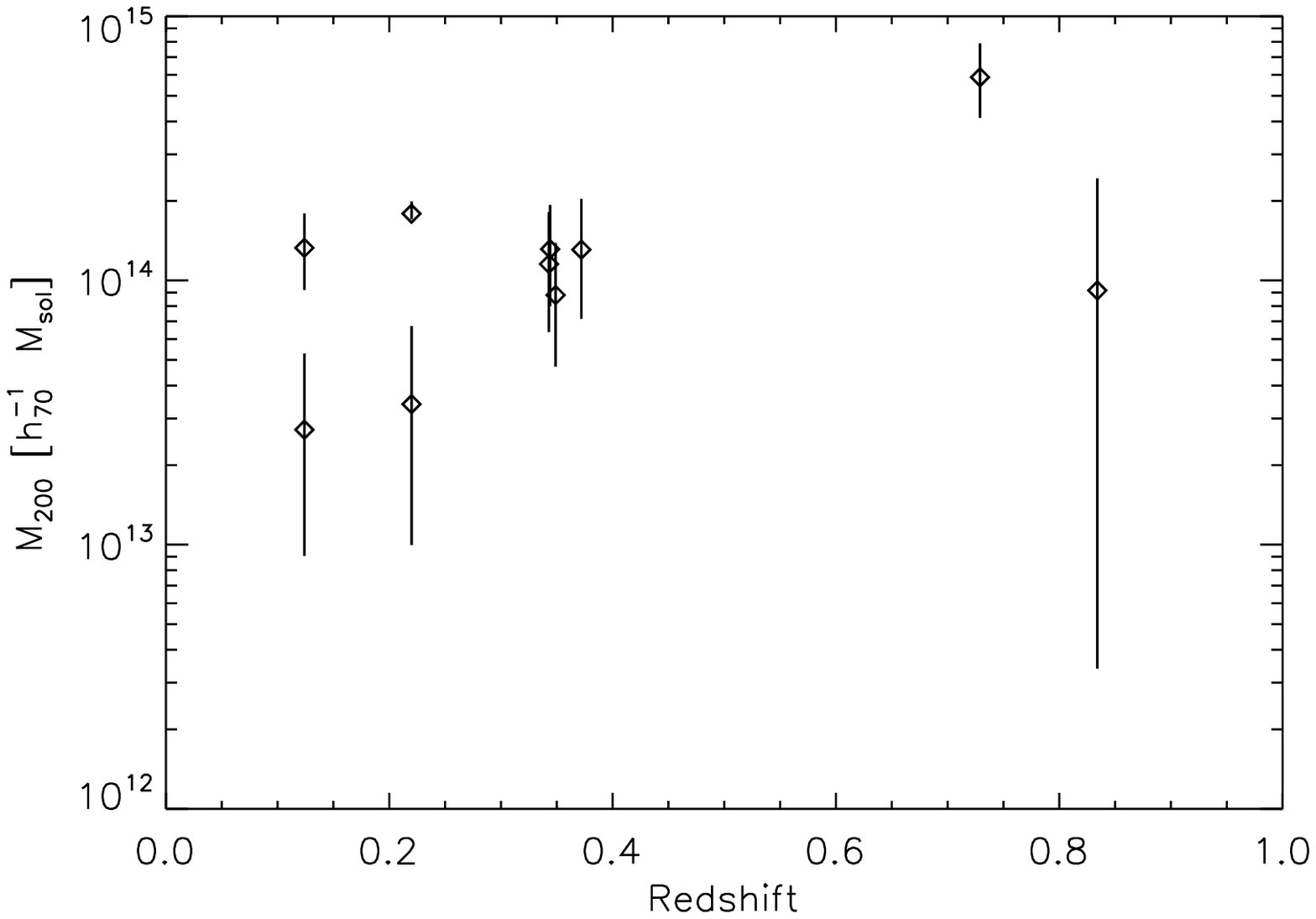}
      \caption{\btxt{Plot showing weak lensing mass M$_{200}$ v.s. redshift $z$ of the COSMOS systems analyzed in this work.}
              }
\label{Fig:m200z}
\end{figure}

\btxt{We estimated the masses by fitting $\Delta \Sigma_{NFW}$ to the measured $\Delta \Sigma$ (Section \ref{sect:wlsign}), in a radial range of 0.1--4 Mpc. The data were not binned for the fit. We used the Metropolis-Hastings Markov Chain Monte Carlo algorithm for $\chi^2$ minimization (see Fig. \ref{Fig:m200chisq} and Fig. \ref{Fig:m200prof}) and found best-fit M$_{200}$ in the range of $\sim$ 0.3--6 $\times 10^{14}$ h$^{-1}_{70}$ M$_{\odot}$ (see Fig. \ref{Fig:m200z} and Table \ref{tab:wl}).} This mass range is consistent with the low X-ray temperatures described above.

\begin{deluxetable*}{ccccccc}
\tablewidth{0pt}
\tablecaption{Results of the weak lensing analysis.\label{tab:wl}} %
\tablehead{
\colhead{id} &\colhead{M$_{500}$ \tablenotemark{a}} & \colhead{M$_{200}$ \tablenotemark{a}} & \colhead{c$_{200}$ \tablenotemark{b}} & \colhead{$\chi^2$ \tablenotemark{c} } & \colhead{degrees of} & \colhead{MMGG / X-ray} \\
\colhead{ } & \colhead{$10^{14}$ h$^{-1}_{70}$ M$_{\odot}$} & \colhead{$10^{14}$ h$^{-1}_{70}$ M$_{\odot}$} & \colhead{ } & \colhead{ } & \colhead{freedom \tablenotemark{d}} & \colhead{centering ratio \tablenotemark{e}} 
}
\startdata
11   & $1.28^{+0.14}_{-0.06}$ & $1.79^{+0.20}_{-0.09}$ & 4.74 & 25762.57 & 22571 & $1.03^{+0.39}_{-0.34}$ \\
17   & $0.92^{+0.52}_{-0.42}$ & $1.31^{+0.73}_{-0.59}$ & 4.38 & 12749.11 & 10960 & $1.00^{+0.56}_{-0.45}$ \\
25   & $0.20^{+0.19}_{-0.14}$ & $0.27^{+0.26}_{-0.18}$ & 6.42 & 73753.62 & 64811 & $1.00^{+0.94}_{-0.67}$ \\
29   & $0.93^{+0.44}_{-0.36}$ & $1.31^{+0.62}_{-0.51}$ & 4.48 & 19686.50 & 16968 & $0.99^{+0.47}_{-0.39}$ \\
120  & $0.60^{+1.00}_{-0.58}$ & $0.92^{+1.51}_{-0.88}$ & 3.22 & 5122.80 & 4296 &   $1.00^{+1.65}_{-0.96}$ \\
149  & $0.97^{+0.34}_{-0.30}$ & $1.33^{+0.47}_{-0.41}$ & 5.38 & 103367.55 & 91433 & $0.99^{+0.163}_{-0.32}$ \\
193  & $0.25^{+0.25}_{-0.18}$ & $0.34^{+0.33}_{-0.24}$ & 5.75 & 47237.02 & 41059 & $1.01^{+0.98}_{-0.71}$ \\
220  & $3.76^{+1.29}_{-1.12}$ & $5.88^{+2.01}_{-1.75}$ & 2.85 & 7443.86 & 6108 & $0.80^{+0.33}_{-0.28}$ \\
237  & $0.63̂^{+0.36}_{-0.29}$ & $0.88^{+0.50}_{-0.41}$ & 4.70 & 21859.89 & 19021 & $1.06^{+0.47}_{-0.50}$ \\
262  & $0.82^{+0.47}_{-0.37}$ & $1.15^{+0.66}_{-0.52}$ & 4.54 & 10039.91 & 8546 & $1.01^{+0.57}_{-0.45}$ \\
\enddata
\tablenotetext{a}{centered on the X-ray peak}
\tablenotetext{b}{halo concentration of the best-fit NFW profile given by the mass-concentration relation in Eq \ref{eq:Mc}}
\tablenotetext{c}{$\chi^2$ of the best-fit model}
\tablenotetext{d}{the number of source galaxies in the weak lensing analysis for each system is given by the degrees of freedom + 1}
\tablenotetext{e}{the ratio of M$_{200}$ centered on the MMGG to M$_{200}$ centered on the X-ray peak, see Section \ref{sect:cent}}
\end{deluxetable*}

\subsection{Centering comparison}
\label{sect:cent}
\citet{2012ApJ...757....2G} \btxt{\citep[see also][]{2011ApJ...726...48H}} showed that miscentering the dark matter halo can bias the lensing mass of the halo low. Therefore we investigated the effects of the uncertainty of the centering of the dark matter halo on our lensing mass estimates by performing the weak lensing analysis described above with centering  on the locations of the X-ray peaks and MMGGs \btxt{\citep[from][]{2011ApJ...742..125G}}, and comparing the resulting halo masses.  

\begin{figure}[htb]
\epsscale{1.0} 
\plotone{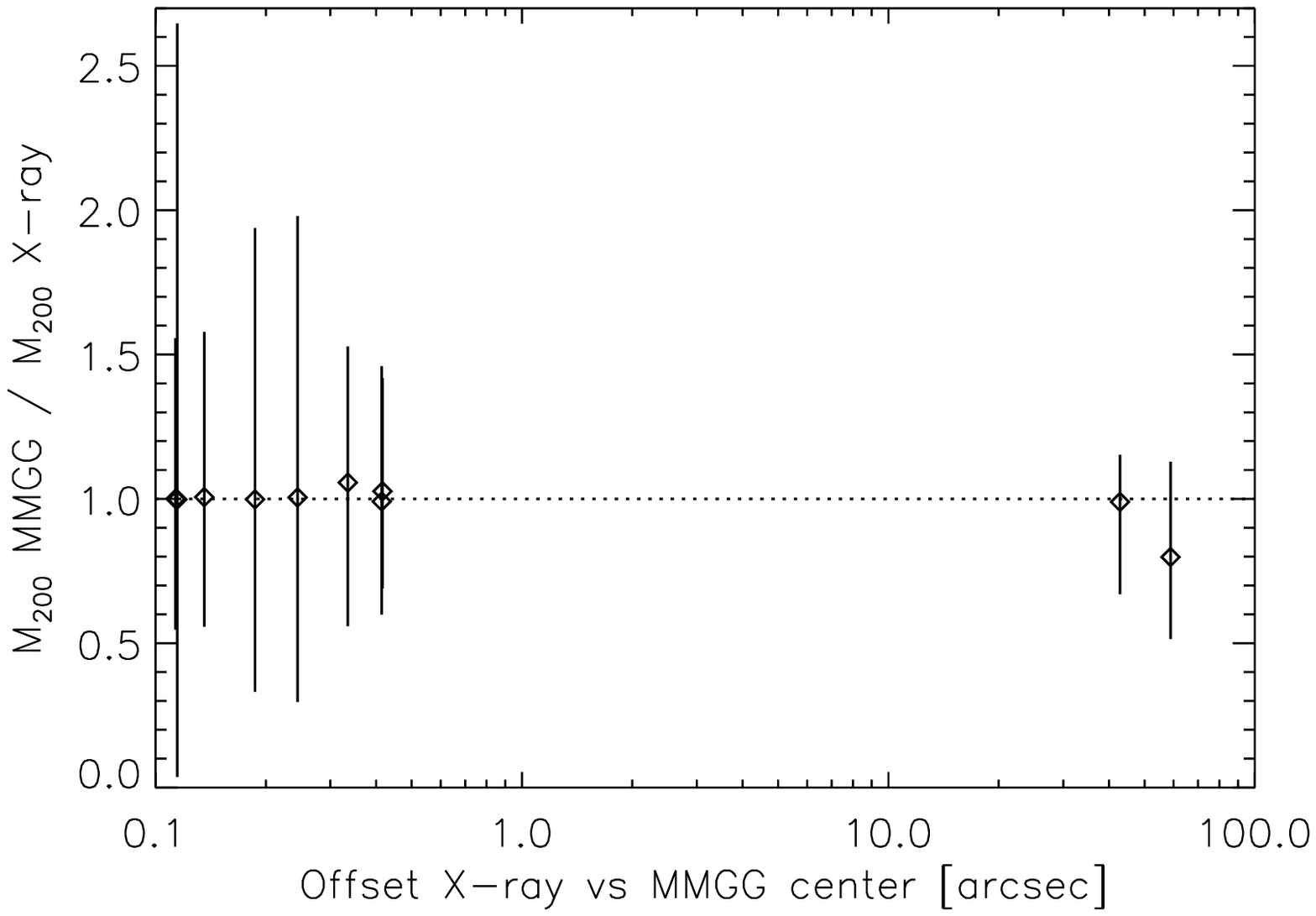}
      \caption{\btxt{Plot showing the ratio of lensing mass estimates for COSMOS galaxy groups centering on the most massive group galaxies (MMGGs) to mass estimates centering on the locations of the X-ray peaks v.s. the offset between the location of MMGGs and X-ray peaks.}
              }
\label{Fig:cent}
\end{figure}

\btxt{The offset between the MMGGs and X-ray peaks are typically less than the uncertainty of the position of the X-ray centroid, which is given by 32 arcsec divided by the signal-to-noise ratio ($\sim$ 10--15 for our sample) for XFLAG=1 groups in the COSMOS group catalog (see Fig. \ref{Fig:cent}). The only exceptions are groups with X-ray id\# 149 and 220, which have an offset of 43 and 59 arcsec respectively.
}

\btxt{The best-fit M$_{200}$ using MMGG and X-ray centering are typically consistent within a few per cent (Table \ref{tab:wl} and Fig. \ref{Fig:cent}). The only deviant group is X-ray id\# 220, which has a MMGG centered mass $\sim$ 20\% lower than the X-ray centered. This system has a peculiar S-shape morphology, which makes accurate center determination difficult \citep{2007ApJS..172..254G}. However, the mass discrepancy with MMGG and X-ray centering is at a less than 1$\sigma$ statistical significance (see also Section \ref{sect:id220} for further discussion on the this system). 
}  

\btxt{A miscentered cluster is expected to show a suppression in the lensing signal at small scales. We do not detect this effect in the mass surface density contrast profiles (Fig. \ref{Fig:m200prof}), including the two groups with significant offsets between MMGG and X-ray centers. We thus conclude that the chosen X-ray centers are accurate and that our lensing masses are not significantly affected by uncertainties in centering. 
}

\subsection{Bias due to M--c relation}
\label{sect:Mc}
A possible systematic in the lensing analysis is an incorrect assumed mass -- concentration relation for the NFW profile (Eq. \ref{eq:Mc}). \btxt{E.g. \citet{2012MNRAS.427.1298H} showed that varying the normalisation of the M--c relation by $\pm$ 20 \% biases lensing NFW mass estimates by $\sim$ 5--15 \%, depending on the mass definition. However, the sensitivity of NFW mass estimates to possible biases in the M--c relation diminishes when the mass estimates are extended further from the cluster center.}

\btxt{Our lensing masses are measured within R$_{200}$ and they are} consistent with the stacked lensing analysis of galaxy groups in the COSMOS field by  \citet{2010ApJ...709...97L}, who used the M-c relation of \citet{2009ApJ...707..354Z} instead of the \citet{2008MNRAS.390L..64D} relation used by us. Furthermore, the mass range implied by both our lensing analysis and the lensing analysis of \citet{2010ApJ...709...97L} is consistent with the typical dark matter halo mass derived with clustering analysis in the COSMOS field \citep{2012ApJ...758...47A}. An incorrect assumed NFW concentration would result in lensing masses contradicting the clustering analysis.

\subsection{Massive galaxy group at $z$ = 0.73}
\label{sect:id220}
\citet{2007ApJS..172..254G} performed a weak lensing analysis of the massive galaxy group at redshift z = 0.73 in the COSMOS field, with id \#220 in the X-ray group catalog. They reported a very high weak lensing mass of $6 \pm 3 \times 10^{15}$ \Msol for the dark matter halo, which is in apparent tension with the X-ray mass M$_{500} \simeq 1.6 \times 10^{14}$ \Msol derived from their X-ray spectroscopic temperature T$_X = 3.51^{+0.60}_{-0.46}$ keV using M-T relations from the literature.

Our X-ray spectroscopic temperature of 4.6$^{+1.0}_{-0.7}$ keV is consistent with the X-ray analysis of \citet{2007ApJS..172..254G}. However, we found a weak lensing \btxt{M$_{200}$ of 4.12$^{+1.41}_{-1.23} \times 10^{14}$ \Msol} \citep[scaled to $h = 1.0$ as used by][]{2007ApJS..172..254G}. This is over an order of magnitude lower than the lensing mass of \citet{2007ApJS..172..254G}, but consistent within errors with the mass predictions from X-ray analyses. This implies that the previously reported high lensing mass is the total mass of the whole superstructure, whereas the lower mass implied by both X-rays and our lensing analysis is the mass of the galaxy group. This argument is further supported by the clustering analysis of groups in the COSMOS field \citep[see Section \ref{sect:Mc} and][]{2012ApJ...758...47A}. \btxt{We further note that exclusion of this source from our sample would not affect our results.}

\section{M--T scaling relation}
\label{sect:scal}
We used our center excised X-ray temperatures and weak lensing group masses in the COSMOS field (Table \ref{tab:x-ray} and \ref{tab:wl}) to calibrate the scaling relation between these two quantities. As the systems in our sample have both low mass and temperature, we are probing a largely unexplored region of the mass -- temperature plane.

In the self-similar model cluster and group mass and temperature are related by a power-law
\begin{equation}
\label{eq:mtpow}
M \times E(z) = N \times T_X^{\alpha}.
\end{equation}
with slope $\alpha = 3 / 2$ \citep{1986MNRAS.222..323K}. Here $E(z)$, defined as
\begin{equation}
\label{eq:ez}
E(z) = \frac{H(z)}{H_0} = \sqrt{\Omega_M ~(1 + z)^3 + \Omega_{\lambda}}
\end{equation}
for flat cosmologies, describes the scaling of overdensity with redshift. 

Scaling relations at galaxy group masses are typically derived for M$_{500}$ \citep[e.g.][]{2001A&A...368..749F,2009ApJ...693.1142S,2011A&A...535A.105E}, i.e. the mass inside the radius where the average density is 500 times the critical density of the Universe. We rescaled the lensing masses derived above to this value using the best-fit NFW profiles to enable direct comparison. We assumed the power-law relation given by Eq. \ref{eq:mtpow} and linearised it by taking a logarithm
\begin{equation}
\label{eq:mtrel}
\log_{10} \frac{M_{500} E(z)}{10^{14} h^{-1}_{70}} = \log_{10} N + \alpha \times \log_{10} \frac{T_X}{3 keV}.
\end{equation}
We evaluated the logarithm of the normalization and the slope of the M -- T relation using the FITEXY linear regression method, \btxt{with bootstrap resampling to compute statistical uncertainties of the fit parameters}. 

\begin{figure*}[htb]
 \epsscale{.75}
\plotone{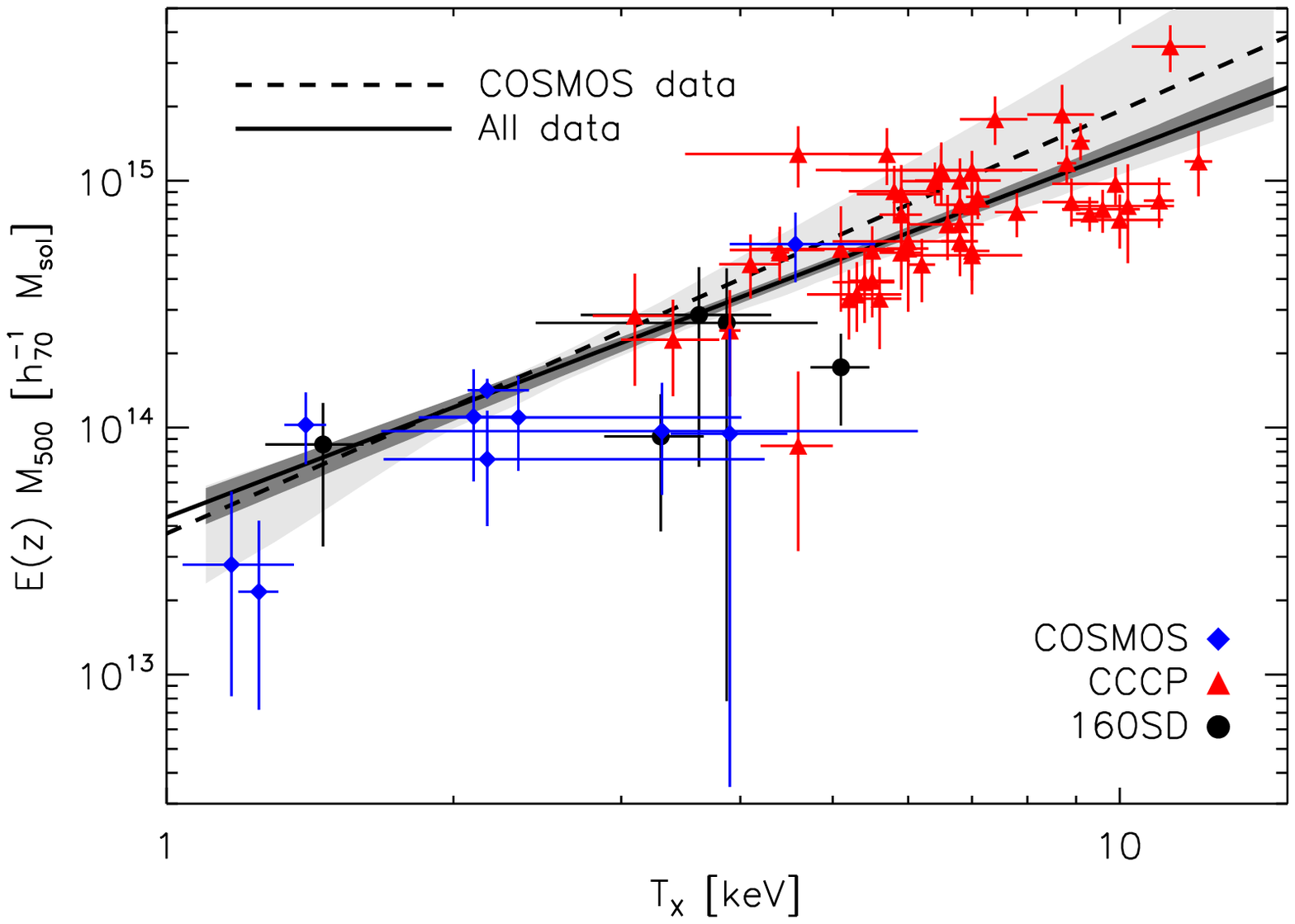}
      \caption{\btxt{The scaling of cluster mass M$_{500}$ to X-ray temperature $T_X$. Blue diamonds show COSMOS systems analyzed in this work, red triangles are systems from the CCCP cluster catalog and black circles are from the 160SD survey. The solid line and dark shaded region shows the best-fit scaling relation with statistical uncertainties fitted to all data points, the dashed line and light shaded region shows the relation fitted to COSMOS data points. }
          }
\label{Fig:MTrel}
\end{figure*}

\begin{figure}[tb]
\epsscale{1.0}
\plotone{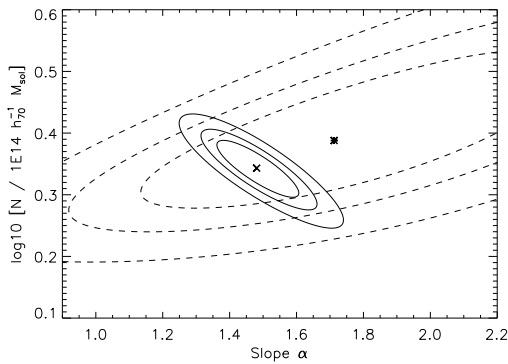}
      \caption{\btxt{Likelihood contours at 68, 90 and 99 \% statistical significance for the parameters of the M$_{500}$ -- $T_X$ scaling relation fitted to COSMOS systems described in this work (dashed contours) and to all data points shown in Fig. \ref{Fig:MTrel} (solid contours). 
              }}
\label{Fig:MTrelCont}
\end{figure}

For the COSMOS systems, we obtained the best-fit parameters $\alpha = 1.71^{+0.57}_{-0.40}$ and $\log_{10} N = 0.39^{+0.04}_{-0.10}$, with $\chi^2 = 5.07$ for 8 degrees of freedom (see Table \ref{tab:mtrel}, Fig. \ref{Fig:MTrel} and \ref{Fig:MTrelCont}). However, as all our systems have low masses and large errors, the constraint on the scaling relation suffers from rather large uncertainties.  

\btxt{We therefore extended our sample with additional measurements at higher temperatures/masses. \citet{2011ApJ...726...48H} determined weak lensing masses for a sample of 25 moderate X-ray luminosity clusters drawn from the 160 square degree survey \citep[160SD][]{1998ApJ...498L..21V,2003ApJ...594..154M}, using HST ACS observations. Unfortunately X-ray temperatures are available for only 5 systems, which we use here. To extend the mass range further we also include measurements for 50 massive clusters that were studied as part of the Canadian Cluster Comparison Project (CCCP). The lensing masses, based on deep CFHT imaging data, are presented in \citet{2012MNRAS.427.1298H}, whereas the X-ray temperatures are taken from \citet{2013ApJ...767..116M}. The X-ray temperatures in \citet{2013ApJ...767..116M} are obtained with both Chandra and XMM-Newton, but Chandra temperatures are adjusted to match XMM-Newton calibration. } 

\btxt{This gives us a total sample of 65 systems with masses and temperatures spanning the range of a few times $10^{13}$ to a few times $10^{15}$ \Msol~ and 1--12 keV.} Fitting the M$_{500}$ -- T$_X$ relation to the whole extended sample we obtained the best-fit parameters $\alpha = 1.48^{+0.13}_{-0.09}$ and $\log_{10} N = 0.34^{+0.02}_{-0.04}$ with $\chi^2 = 112.57$ for 63 degrees of freedom (see Table \ref{tab:mtrel}, Fig. \ref{Fig:MTrel} and \ref{Fig:MTrelCont}). 

\btxt{We evaluated intrinsic scatter of the relation by making a distribution of the ratio of data to best-fit model for each point and computing the dispersion. The resulting scatter in mass at fixed T for the relation fitted to COSMOS data points and to the full sample are consistent, 28 $\pm$ 13 \% and 28 $\pm$ 7 \% respectively, indicating that the samples are consistent with each other. 
}

\begin{deluxetable*}{lccccc}
\tablewidth{0pt}
\tablecolumns{6}
\tablecaption{Best-fit parameters of the M$_{500}$ -- T$_X$ scaling relation.\label{tab:mtrel}}%
\tablehead{ 
\colhead{Sample} & \colhead{Slope} & \colhead{Normalisation} & \colhead{Intrinsic scatter}  & \colhead{$\chi^2$} & \colhead{degrees of} \\
\colhead{ } & \colhead{$\alpha$} & \colhead{$\log_{10} N$} & \colhead{\%}  & \colhead{ } & \colhead{freedom}  
}
 \startdata
COSMOS                      		& $1.71^{+0.57}_{-0.40}$ & $0.39^{+0.04}_{-0.10}$ & 28 $\pm$ 13 & 5.07   & 8  \\
COSMOS+CCCP+160SD					& $1.48^{+0.13}_{-0.09}$ & $0.34^{+0.02}_{-0.04}$ & 28 $\pm$ 7  & 112.57 & 63 \\
COSMOS+CCCP+160SD, modified T$_{X}$	& $1.40^{+0.12}_{-0.10}$ & $0.32^{+0.02}_{-0.03}$ & 35 $\pm$ 9  & 117.99 & 63 \\
 \enddata
\end{deluxetable*}

\section{Discussion}
\label{sect:disc}

\begin{figure*}[htb]
\epsscale{.75}
\plotone{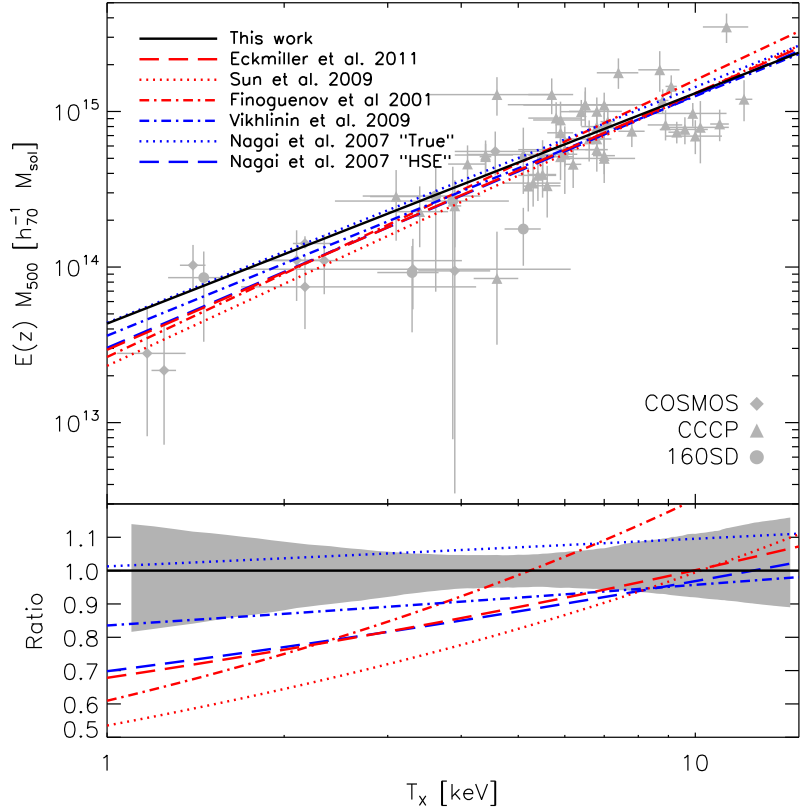}
      \caption{\btxt{{\it Top panel:} Comparison of M$_{500}$ -- T$_X$ relations discussed in the text. The solid line corresponds to the best-fit weak lensing calibrated relation combining COSMOS, CCCP and 160SD samples in this work, data points are shown in gray.
{\it Bottom panel:} Ratio of M$_{500}$ -- T$_X$ relations shown in the top panel to the best-fit relation in this work (solid line). Gray shaded region shows the relative statistical uncertainty of our relation. 
               }}
         \label{Fig:MTrelComp}
   \end{figure*}

 \begin{figure}[htb]
 \epsscale{1.0}
\plotone{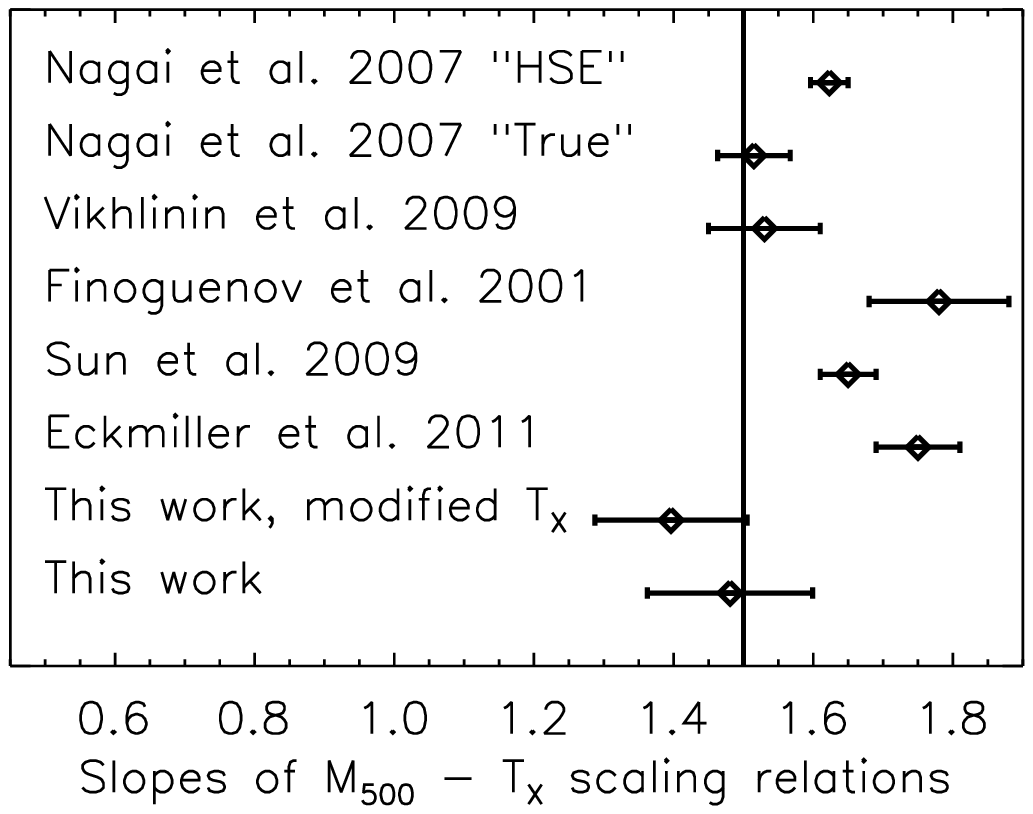}
      \caption{\btxt{Comparison of the slopes of M$_{500}$ -- T$_X$ relations shown in Fig. \ref{Fig:MTrelComp} and Fig. \ref{Fig:MTrelmod}. The vertical line corresponds to the self-similar slope of $3/2$, the error bars describe the 68 \% statistical uncertainties of the slopes.
              }}
         \label{Fig:MTrelSlope}
   \end{figure}  
   
The slope of our best-fit relation of the \btxt{full} sample $1.48^{+0.13}_{-0.09}$ is consistent with the self-similar prediction of 3/2 \citep{1986MNRAS.222..323K}. Unfortunately direct comparison of our best-fit relation to \btxt{most} other weak lensing calibrated M--T relations is not possible. \citet{2010ApJ...721..875O} calibrated deprojected center excised temperatures (whereas our temperatures are projected) to M$_{500}$ for the LoCuSS cluster sample, consisting of only cluster mass systems, and attained a slope of 1.49 $\pm$ 0.58. \citet{2007MNRAS.379..317H} and \citet{2011ApJ...737...59J} calibrated X-ray temperatures to weak lensing M$_{2500}$ for cluster mass systems and attained slopes of 1.34$^{+0.30}_{-0.28}$ and 1.54 $\pm$ 0.23 respectively. As their mass definition differs from ours and masses are thus derived from a smaller region, their relations are not directly comparable to our analysis. In the case of \citet{2011ApJ...737...59J} the clusters are also at a significantly higher redshift than our sample, representing a cluster population at an earlier evolutionary stage. 

\btxt{However, \citet{2013ApJ...767..116M} used the 50 CCCP clusters, which are also included in our sample, to fit scaling relations between X-ray observables and lensing masses. For M$_{500}$--T$_X$ scaling they obtained a slope of 1.97 $\pm$ 0.89 and 1.42 $\pm$ 0.19 with a scatter in mass of 46 $\pm$ 23 \% and 17 $\pm$ 8, using R$_{500}$ dervived from weak lensing and X-ray analysis respectively. Both of these are consistent within the errorbars with our findings.}

\btxt{The fact that the published lensing calibrated M--T relations at cluster masses and our group mass predict consistent slopes, indicates that both clusters and groups follow the same mass-to-temperature scaling. This is in apparent tension with relations relying on HSE mass estimates, which generally predict steeper slopes and lower normalisation when group mass systems are included (see Fig \ref{Fig:MTrelComp}). E.g.} \citet{2001A&A...368..749F} used ASCA observations of the extended HIFLUGCS sample consisting of 88 systems spanning a similar mass and temperature range as our \btxt{full} sample and obtained a slope of 1.636 $\pm$ 0.044 for the M$_{500}$ -- T$_X$ relation, \citet{2009ApJ...693.1142S} calibrated a similar relation to archival Chandra observations of 43 groups and 14 clusters and obtained a slope of 1.65 $\pm$ 0.04, and \citet{2011A&A...535A.105E} obtained a slope of 1.75 $\pm$ 0.06 for a sample consisting of 112 groups and HIFLUGCS clusters. However, e.g. \citet{2009ApJ...692.1033V} used a sample of clusters with T$_X$ $\gtrsim$ 2.5 keV to calibrate a M$_{500}$--T$_X$ relation under the assumption of HSE, and obtained a slope of 1.53 $\pm$ 0.08, consistent with our weak lensing relations.

\btxt{The difference in slope between hydrostatic and our weak lensing calibrated M--T relation is significant at $\sim$ 1--2$\sigma$ level (see Fig. \ref{Fig:MTrelSlope}). The steeper slope and lower normalisation of HSE relations amounts to a temperature dependent bias between the scaling relations at an up to $\sim 2\sigma$ significance (see Fig. \ref{Fig:MTrelComp}, lower panel).} 

\btxt{Simulations indicate that HSE masses may be biased low due to non-thermal pressure support and kinetic pressure from gas motion \citep[e.g.][]{2007ApJ...668....1N,2010ApJ...725.1452S,2012NJPh...14e5018R}. Furthermore, the deviation from self-similarity in the M--T relation implied by HSE mass estimates is hard to reproduce in simulations \citep{2004MNRAS.348.1078B}. Thus the preferred interpretation is a deviation between hydrostatic and lensing masses, amounting to $\sim$ 30--50 \% at 1 keV. Our study provides the first observational support for this scenario at group scales. This effect has previously been observed at cluster masses by \citet{2008MNRAS.384.1567M} and \citet{2013ApJ...767..116M}.
}

The effect of deviation between hydrostatic and lensing masses on scaling relations has previously been studied by \citet{2007ApJ...668....1N}. They simulated a sample of groups and clusters in a mass range approximately consistent with our extended sample, including effects of cooling and star formation. The simulated clusters were used for mock Chandra observations to calibrate an M$_{500}$--T$_X$ relation using both true masses and masses derived under the hydrostatic equilibrium condition. Their best-fit relation using true masses is consistent with our lensing relation whereas their hydrostatic relation very accurately follows the observed hydrostatic relation of \citet{2009ApJ...693.1142S} (see Fig \ref{Fig:MTrelComp} and \ref{Fig:MTrelSlope}). This \btxt{provides further evidence that a bias in hydrostatic masses can affect the shape of scaling relations. }


\subsection{X-ray cross-calibration}
\label{sect:crosscal}
\btxt{Cross-calibration issues in the energy dependence of the effective area of X-ray detectors affects cluster spectroscopic temperatures obtained with different instruments \citep[e.g.][]{2008A&A...478..615S,2010A&A...523A..22N,2013A&A...552A..47K,2013ApJ...767..116M}. Recent observations indicate cluster temperatures measured with Chandra are typically $\sim$ 15 \% higher than those measured with XMM-Newton \citep{2010A&A...523A..22N,2013ApJ...767..116M}. As we compare our lensing calibrated M-T relation relying on XMM-Newton temperature measurements (or Chandra temperatures modified to match XMM-Newton) to Chandra based relations in literature, we investigate here if the detected discrepancies can be attributed to X-ray cross-calibration uncertainties. 
}

\begin{figure}[htb]
 \epsscale{1.0}
\plotone{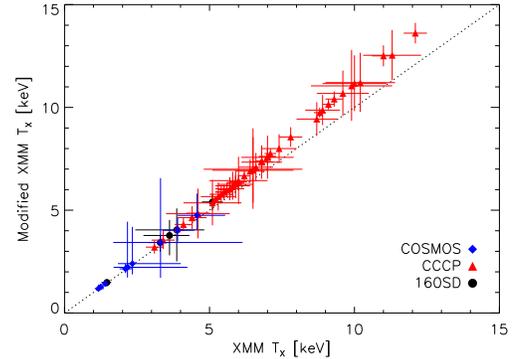}
      \caption{\btxt{Plot showing XMM-Newton X-ray temperatures modified for Chandra calibration v.s. unmodified XMM-Newton temperatures for our group and cluster sample.
              }}
         \label{Fig:Tmod}
   \end{figure}

\btxt{Whereas cluster temperatures $\gtrsim$ 4 keV are typically inferred from the shape of the bremsstrahlung continuun which depends strongly on the energy dependence of the effective area, lower group temperatures are mainly determined from emission lines and are thus independent of energy dependent cross-calibration. This effect is seen in comparisons of group and cluster temperatures obtained with XMM-Newton and Chandra \citep{2008A&A...478..615S}. As the measured energy of a photon at the detector also depends on the redshift of the source, we use the temperature and redshift dependent modification given by
\begin{equation}
T_X^{\rm modified} = T_X^{\rm XMM} \times \Big(1 + \frac{0.15 ~ T_X^{\rm XMM}}{\rm 10 keV} \frac{1}{1 + z}\Big)
\end{equation}
to modify our XMM-Newton based temperatures to match Chandra calibration (see Fig \ref{Fig:Tmod}). 
}

\begin{figure*}[htb]
 \epsscale{0.75}
\plotone{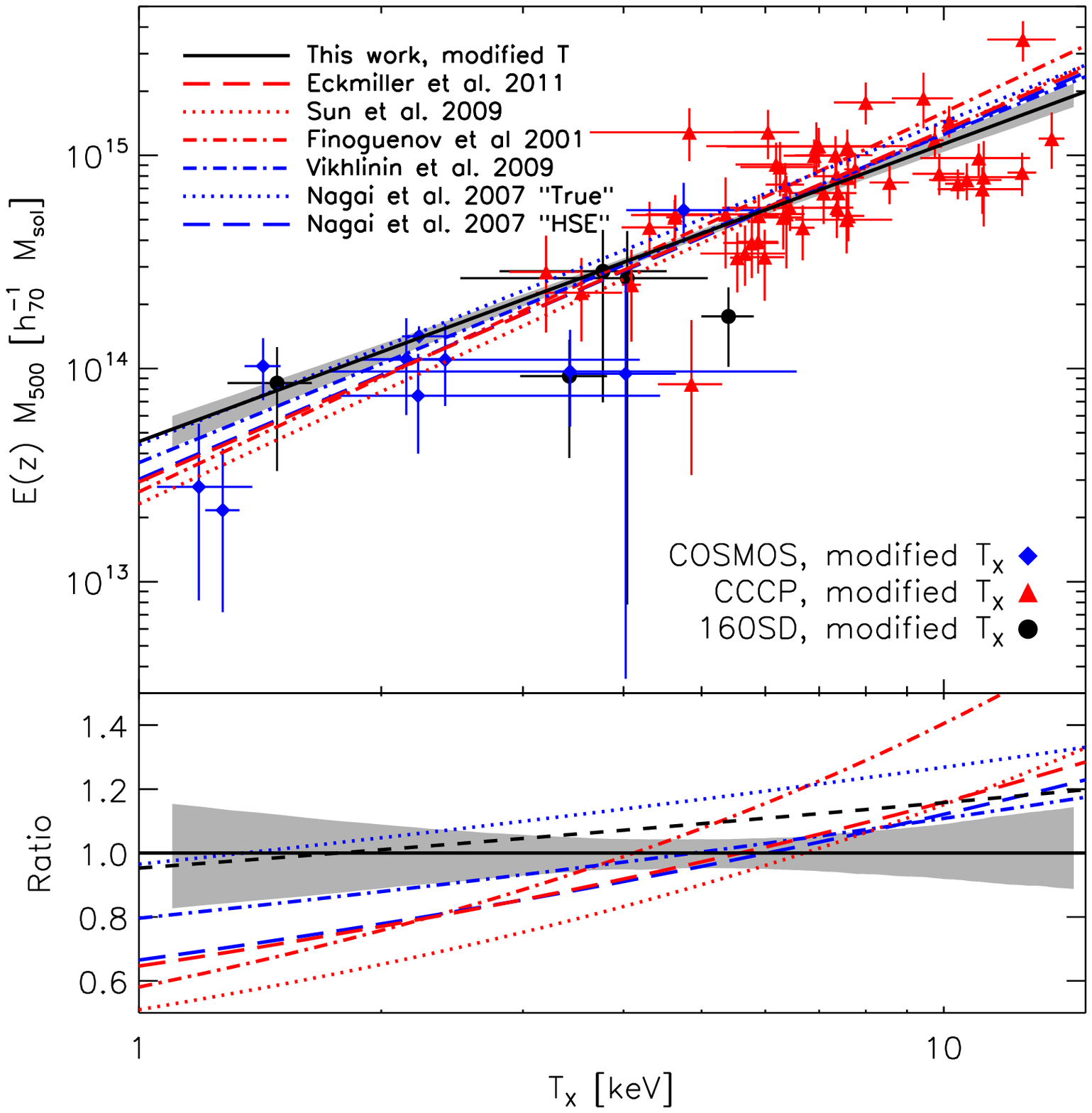}
      \caption{\btxt{{\it Top panel:} The solid line and shaded region showns M$_{500}$ -- T$_X$ relation and statistical uncertainties using the XMM-Newton temperatures modified for Chandra calibration, with comparison to other relations discussed in the text.
	{\it Bottom panel:} Ratio of the relations shown in the top panel to the relation fitted  data with XMM-Newton temperatures modified for Chandra calibration (solid line). The shaded region shows the relative statistical uncertainty of the modified XMM relation.   
              }}
         \label{Fig:MTrelmod}
   \end{figure*}

\btxt{Re-fitting the M$_{500}$--T$_{X}$ relation with the modified XMM temperatures, we find a marginally flatter slope than using unmodified temperatures. The slope still consistent with the self-similar prediction of 3/2 (Table \ref{tab:mtrel}, and Fig. \ref{Fig:MTrelSlope} and \ref{Fig:MTrelmod}). Comparing to Chandra based HSE relations from literature, we find that HSE still predicts lower masses at group scales than lensing. We conclude that the differences between HSE and lensing M-T relations can not be explained by X-ray cross-calibration uncertainties and that lensing calibrated relations have slopes consistent with self-similarity for both Chandra and XMM-Newton based temperatures. 
}

\section{Summary and conclusions}
\label{sect:sum}
\btxt{We calibrated a scaling relation between weak lensing masses and spectroscopic X-ray temperatures for a sample of 10 galaxy groups in the COSMOS field, 5 clusters from the 160SD survey, and 50 clusters from the CCCP survey. This gave a sample of 65 systems, spanning a wide mass and temperature range of $M_{500} \sim 10^{13}$--$10^{15}$ \msun~ and $T_X \sim$1--12 keV, extending weak lensing calibrated M--T relations to an unexplored region of the mass -- temperature plane.  }

\btxt{We found that the best-fit slope of the relation is consistent with the prediction for self-similar cluster evolution of \citet{1986MNRAS.222..323K}. This is in apparent tension with M--T relations at group scales in literature, which use X-ray masses derived under HSE. These relations typically predict steeper slopes and lower normalizations. }

\btxt{The deviations from self-similarity implied by HSE relations are likely due to HSE masses being biased low  in comparison to unbiased lensing masses. We find that the bias increases with decreasing temperature, amounting to $\sim$30--50 \% at 1 keV. This effect has been detected in simulations and our study provides the first observational evidence for it at group scales. We also show that this effect is not a product of cross-calibration issues between X-ray detectors.}

\btxt{We conclude that this work demonstrates the importance of unbiased weak lensing calibrated scaling relations for precision cosmology with galaxy clusters and groups. Although costly, more weak lensing surveys of galaxy groups are needed to extend the statistical analysis of this work.
}
  
 \acknowledgments
 The authors thank F. Miniati for useful discussion. KK acknowledges support from the Magnus Ehrnrooth Foundation. \btxt{AF acknowledges the Academy of Finland (decision 266918)}. RM is supported by a Royal Society University Research Fellowship and ERC grant MIRG-CT-208994. JR was supported by JPL, which is run by Caltech under a contract for NASA. \btxt{HH acknowledges NWO Vidi grant 639.042.814.} This research has made use of NASA's Astrophysics Data System.

 {\it Facilities:} \facility{HST (ACS)},  \facility{XMM (EPIC)}.  

    \end{document}